\documentclass[11pt,twoside]{article}

\usepackage{asp2006}
\usepackage{graphicx}

\begin{document}
\title{Metal-rich debris discs around white dwarfs}
\author{B.T. G\"ansicke, T.R. Marsh, J. Southworth, A. Rebassa-Mansergas}
\affil{University of Warwick, Dept. of Physics, Coventry CV4~7AL, UK}

\begin{abstract}
We have identified two moderately hot ($\sim$18000--22000\,K) white
dwarfs, SDSS\,J1228+1040 and SDSS\,J1043+0855, which exhibit
double-peaked emission lines in the CaII\,$\lambda\lambda$\,8600
triplet. These line profiles are unambiguous signatures of gaseous
discs with outer radii of $\sim1R_\odot$ orbiting the two white
dwarfs. Both stars accrete from the circumstellar material, resulting
in large photospheric Mg abundances. The absence of hydrogen emission
from the discs, and helium absorption in the white dwarf photospheres
demonstrates that the circumstellar material is depleted in volatile
elements, and the most likely origin of these gaseous rings are
tidally disrupted rocky asteroids. The relatively high mass of
SDSS\,J1228+1040 implies that planetary systems  can not only form
around $4-5\,M_\odot$ stars, but may also survive their post
main-sequence evolution.
\end{abstract}

\section{Introduction}
While more than 250 extra-solar planets orbiting main-sequence stars
have been discovered, the destiny of planetary systems in the late
stages of the evolution of their host stars is very uncertain, and so
far no planet has been found around a white dwarf. Infrared excess
detected around a number of white dwarfs has been interpreted as the
signature of dust discs \citep[e.g.][]{zuckerman+becklin87-1,
becklinetal05-1, kilicetal06-1, vonhippeletal07-1}. The photospheres
of these white dwarfs are rich in metals \citep{zuckermanetal07-1},
indicating ongoing accretion from the circumstellar material.  The
likely origin of these debris discs are tidally disrupted asteroids
\citep{jura03-1}, and hence they represent a close association with
the planetary systems that the white dwarf progenitor stars may have
had. However, while the infrared excess detected around these white
dwarfs can be explained in terms of a dusty debris disc, the
observations actually do not provide any strong constraint on the
geometry of the source of the infrared light
\citep[e.g.][]{reachetal05-1}.  We summarise here our recent discovery
of two white dwarfs in the SDSS spectroscopic data base which exhibit
double-peaked emission lines of Ca\,II$\lambda\lambda$8600,
unambiguously confirming a circumstellar disc-like structure
\citep{gaensickeetal06-3, gaensickeetal07-1}

\section{Why discs? Why metal-rich? Why planetary debris?}
Visually inspecting the SDSS spectra of several hundred white dwarfs,
we noticed very unusual Ca\,II$\lambda\lambda$8600 emission lines in
SDSS\,J122859.93+104032.9 (Fig.\,1) and SDSS\,J104341.53+085558.2
(Fig.\,2). The double-peaked shape of these line profiles is the
unmistakable signature of gas rotating around these stars on Keplerian orbits
\citep{horne+marsh86-1, littlefairetal06-2}.

\begin{figure}
\centerline{
\includegraphics[width=0.83\textwidth]{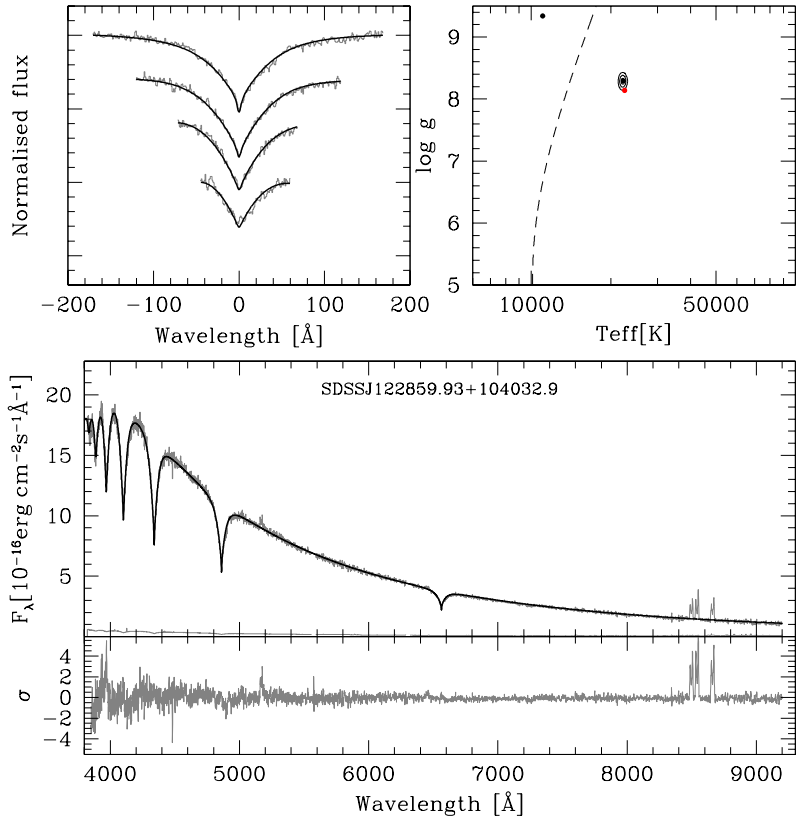}}
\caption{The Sloan spectrum of SDSS\,J1228+1040 along with white dwarf
  model fits. Top left: normalised H$\beta$ to H$\epsilon$ line
  profiles (top to bottom, gray lines) along with the best-fit white
  dwarf model (black lines). Top right panel: 1, 2, and 3$\sigma$
  $\chi^2$ contour plots in the $T_\mathrm{eff}-\log g$ plane. The
  line profiles allow a ``hot'' and a ``cold'' solution (black dots)
  on either side of the temperature corresponding to maximum Balmer
  line equivalent widths (dashed line). The degeneracy in the line
  profile fits is lifted by the best-fit to the continuum slope (red
  dot). The best-fit to the line profiles results in
  $T_\mathrm{eff}=22\,292\pm296$\,K and $\log g=8.29\pm0.05$. Bottom
  panels: the white dwarf spectrum and associated flux errors (gray
  lines) along with the best-fit white dwarf model (black line) to
  the 3850--7150\,\AA\ wavelength range (top) and the residuals of the
  fit (gray line, bottom). Note the strong, double-peaked emission
  lines of Ca\,II$\,\lambda\lambda$\,8600. }
\end{figure}

While Ca\,II emission lines are observed in a few white dwarfs with
low-mass companions \citep[e.g.][]{marsh+duck96-1}, these lines are
\textit{always} accompanied by Balmer emission. In fact, even
WD\,0137--349 and SDSS\,1035+0551, two white dwarf/brown dwarf
binaries, display copious amounts of H$\alpha$ emission
\citep{maxtedetal06-1, littlefairetal06-2}. In contrast to those
binaries, the spectrum of SDSS\,J1228+1040 is totally devoid of Balmer
emission. We obtained time-resolved intermediate resolution
spectroscopy and photometric time series of SDSS\,J1228+1040, which
clearly rule out the possibility of this object being a white dwarf
plus low mass companion binary.  Photospheric Mg\,I\,$\lambda$\,4482
absorption lines demonstrates that both SDSS\,J1228+1040 and
SDSS\,J1043+0855 are accreting from their circumstellar discs.
Stellar parameter of the two white dwarfs (Fig.\,1 \& 2, Table\,1)
were determined from spectral fits using TLUSTY and SYNSPEC models
\citep{hubeny+lanz95-1}.
The absence of hydrogen (and/or helium) emission from material close
to a $\sim20\,000$\,K white dwarf implies that the circumstellar
material must be depleted in volatile elements.  A low content
($\la10$\,\% solar) of helium in the circumstellar material is
corroborated by the absence of helium lines in the spectrum of
SDSS\,J1228+1040.
A dynamical model of the Ca\,II line profiles in SDSS\,J1228+1040
computed following \cite{horne+marsh86-1} implies an outer radius of
the disc of just 1.2 solar radii. This circumstellar material can not
have survived the red giant phase of the white dwarf progenitor at its
current location, and must have been supplied from distances greater
than 1000\,$R_\odot$ after the formation of the white dwarf. A
plausible originally is the tidal disruption of a rocky asteroid
\citep{jura03-1}, probably destabilised from its original much wider
orbit by interaction with a larger object \citep{debesetal02-1}.

\begin{figure}
\centerline{
\includegraphics[angle=-90,width=0.86\textwidth]{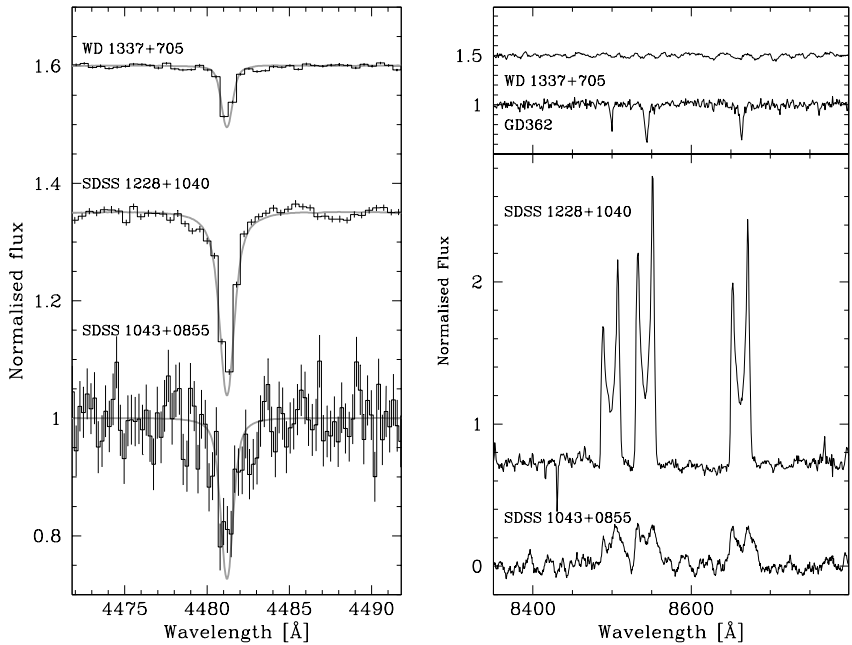}}
\caption{Left panel: The Mg\,I absorption lines in the moderately hot
three DAZ white dwarfs WD\,1337+705, SDSS\,J1228+1040, and
SDSS\,J1043+0855. Best-fit models that were used to determine the
abundances are shown as gray lines. Right panels: The spectra of these
three DAZ plus that of the cold DAZ GD\,362 centred on the
Ca\,II$\lambda\lambda$8600 triplet. The spectra of SDSS\,J1228+1040
and SDSS\,J1043+0855 display double-peaked Ca\,II emission lines
originating in debris discs with outer radii of about 1$R_\odot$. In
GD\,362 photospheric Ca\,II absorption lines are observed, and no
Ca\,II features are present in the spectrum of WD\,1337+705. The
comparison of these four objects suggests that Ca\,II emission can
occur only around white dwarfs which are sufficiently hot to
ionise/excite Ca in the circumstellar debris disc, and that the
equivalent width of the Ca\,II emission may be correlated with the
accretion-induced photospheric metal abundance.}
\end{figure}

\begin{table}[t]
\caption{Stellar parameters of SDSS\,J1228+1040 and SDSS\,J1043+0855.}
\smallskip
\begin{center}{\small
\setlength{\tabcolsep}{0.95ex}
\begin{tabular}{ccccccc}
\tableline
\noalign{\smallskip}
System & $T_\mathrm{eff}$ & $M_\mathrm{WD}$ & $\tau_\mathrm{cool}$ &
$M_\mathrm{MS-prog.}$ & Mg abund. & Ca\,II E.W. \\
   & [K] & $[M_\odot]$ & [yr] & [$M_\odot$] & [$\odot$] & [\AA] \\
\noalign{\smallskip}
\tableline
\noalign{\smallskip}
SDSS\,J1228+1040 & 22292 & 0.81 & $1.0\times10^8$\,yr & $\sim4$  & 0.7
& 61.1 \\
SDSS\,J1043+0855 & 18330 & 0.67 & $1.8\times10^8$\,yr & $\sim2.7$ &
0.3 & 21.2\\
\noalign{\smallskip}
\tableline
\end{tabular}}
\end{center}
\end{table}

\section{Implications and conclusions}
It has been suggested that planetary systems may survive the post-main
sequence evolution of their host stars \citep{burleighetal02-1,
villaver+livio07-1}, however, no planet has yet been discovered around
a white dwarf. The detection of debris discs from rocky asteroids
around white dwarfs, such as SDSS\,J1228+1040 and SDSS\,J1043+0855 and
the cooler white dwarfs with dusty debris discs lends strong support
to the survival hypothesis. It appears also entirely possible that
these white dwarfs may still have planetesimal objects or
planets. SDSS\,J1228+1040 is particular interesting, as its relatively
high mass implies that its progenitor must have had a mass of
$\sim4\,M_\odot$ \citep{dobbieetal06-1}, suggesting that also
short-lived massive stars may be host to planetary discs. This is in
accordance with the detection of a relatively massive debris disc
around the young A2e star MWC\,480 \citep{manningsetal97-1}.



\begin{thebibliography}{}

\bibitem[\protect\astroncite{{Becklin} et~al.}{2005}]{becklinetal05-1}
{Becklin} E.~E., {Farihi} J., {Jura} M., {Song} I., {Weinberger} A.~J.,
  {Zuckerman} B., 2005,
  ApJ Lett.\   632, L119

\bibitem[\protect\astroncite{{Burleigh} et~al.}{2002}]{burleighetal02-1}
{Burleigh} M.~R., {Clarke} F.~J., {Hodgkin} S.~T., 2002,
  MNRAS\   331, L41

\bibitem[\protect\astroncite{{Debes} \& {Sigurdsson}}{2002}]{debesetal02-1}
{Debes} J.~H., {Sigurdsson} S., 2002,
  ApJ\   572, 556

\bibitem[\protect\astroncite{{Dobbie} et~al.}{2006}]{dobbieetal06-1}
{Dobbie} P.~D., {Napiwotzki} R., {Burleigh} M.~R., {Barstow} M.~A., {Boyce}
  D.~D., {Casewell} S.~L., {Jameson} R.~F., {Hubeny} I., {Fontaine} G., 2006,
  MNRAS\   369, 383

\bibitem[\protect\astroncite{{G{\"a}nsicke} et~al.}{2007}]{gaensickeetal07-1}
{G{\"a}nsicke} B.~T., {Marsh} T.~R., {Southworth} J., 2007,
  MNRAS\   380, L35

\bibitem[\protect\astroncite{{G{\"a}nsicke} et~al.}{2006}]{gaensickeetal06-3}
{G{\"a}nsicke} B.~T., {Marsh} T.~R., {Southworth} J., {Rebassa-Mansergas} A.,
  2006,
  Science\   314, 1908

\bibitem[\protect\astroncite{{Horne} \& {Marsh}}{1986}]{horne+marsh86-1}
{Horne} K., {Marsh} T.~R., 1986,
  MNRAS\   218, 761

\bibitem[\protect\astroncite{{Hubeny} \& {Lanz}}{1995}]{hubeny+lanz95-1}
{Hubeny} I., {Lanz} T., 1995,
  ApJ\   439, 875

\bibitem[\protect\astroncite{{Jura}}{2003}]{jura03-1}
{Jura} M., 2003,
  ApJ Lett.\   584, L91

\bibitem[\protect\astroncite{{Kilic} et~al.}{2006}]{kilicetal06-1}
{Kilic} M., {von Hippel} T., {Leggett} S.~K., {Winget} D.~E., 2006,
  ApJ\   646, 474

\bibitem[\protect\astroncite{{Littlefair} et~al.}{2006}]{littlefairetal06-2}
{Littlefair} S.~P., {Dhillon} V.~S., {Marsh} T.~R., {G{\"a}nsicke} B.~T.,
  {Southworth} J., {Watson} C.~A., 2006,
  Science\   314, 1578

\bibitem[\protect\astroncite{{Mannings} et~al.}{1997}]{manningsetal97-1}
{Mannings} V., {Koerner} D.~W., {Sargent} A.~I., 1997,
  \nat\   388, 555

\bibitem[\protect\astroncite{{Marsh} \& {Duck}}{1996}]{marsh+duck96-1}
{Marsh} T.~R., {Duck} S.~R., 1996,
  MNRAS\   278, 565

\bibitem[\protect\astroncite{{Maxted} et~al.}{2006}]{maxtedetal06-1}
{Maxted} P.~F.~L., {Napiwotzki} R., {Dobbie} P.~D., {Burleigh} M.~R., 2006,
  Nat\   442, 543

\bibitem[\protect\astroncite{{Reach} et~al.}{2005}]{reachetal05-1}
{Reach} W.~T., {Kuchner} M.~J., {von Hippel} T., {Burrows} A., {Mullally} F.,
  {Kilic} M., {Winget} D.~E., 2005,
  ApJ Lett.\   635, L161

\bibitem[\protect\astroncite{{Villaver} \& {Livio}}{2007}]{villaver+livio07-1}
{Villaver} E., {Livio} M., 2007,
  \apj\   661, 1192

\bibitem[\protect\astroncite{{von Hippel} et~al.}{2007}]{vonhippeletal07-1}
{von Hippel} T., {Kuchner} M.~J., {Kilic} M., {Mullally} F., {Reach} W.~T.,
  2007,
  ApJ\   662, 544

\bibitem[\protect\astroncite{{Zuckerman} \&
  {Becklin}}{1987}]{zuckerman+becklin87-1}
{Zuckerman} B., {Becklin} E.~E., 1987,
  Nat\   330, 138

\bibitem[\protect\astroncite{{Zuckerman} et~al.}{2007}]{zuckermanetal07-1}
{Zuckerman} B., {Koester} D., {Melis} C., {Hansen} B., {Jura} M., 2007,
  ApJ\   in press, arXiv:0708.0198

\end{thebibliography}

\end{document}